\documentclass[twocolumn, showpacs,prb]{revtex4}
\usepackage{graphicx}
\usepackage{dcolumn}
\usepackage{bm}

\begin{document}

\title{Enhancement of valley susceptibility upon complete spin-polarization}

\date{\today}

\author{Medini\ Padmanabhan}

\author{T.\ Gokmen}

\author{M.\ Shayegan}

\affiliation{Department of Electrical Engineering, Princeton
University, Princeton, NJ 08544}

\begin{abstract}

Measurements on a two-dimensional electron system confined to an
AlAs quantum well reveal that, for a given electron density, the
valley susceptibility, defined as the change in valley population
difference per unit strain, is enhanced as the system makes a
transition from partial to full spin-polarization. This observation
is reminiscent of earlier studies in which the spin susceptibility
of AlAs electrons was observed to be higher in a single-valley
system than its two-valley counterpart.

\end{abstract}


\maketitle

\begin{figure} \includegraphics[scale=1]{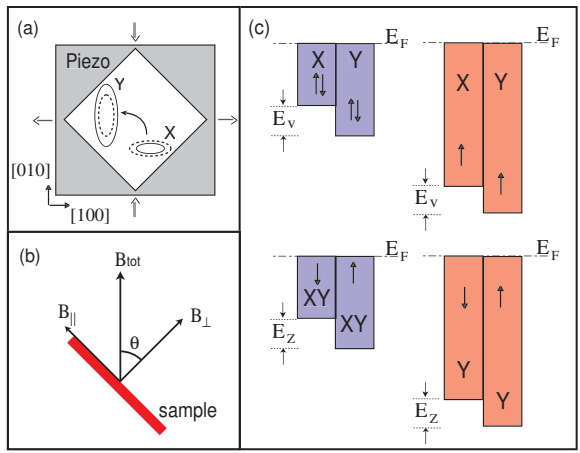}
\caption{ (color online). (a) Schematic showing the strain-induced
transfer of electrons from one valley to another. (b) Orientation of
the sample with respect to applied magnetic field. (c) (top)
Strain-induced valley splitting in two- and single-spin systems.
(bottom) Spin splitting caused by a magnetic field in two- and
single-valley systems.}
\end{figure}

A perennial quest in the study of two dimensional electron systems
(2DESs) has been to understand the role of electron-electron
interaction. Experimental
\cite{okamotoPRL99,shashkinPRL01,pudalovPRL02,zhuPRL03,tutucPRB03,vakiliPRL04,tanPRB06}
and theoretical \cite{attaccalitePRL02,zhangPRB05,depaloPRL05}
reports of enhanced spin susceptibility ($\chi_{s}$) in dilute
systems have indeed provided solid evidence for the increasing
influence of interaction at lower densities where the ratio of
Coulomb to kinetic (Fermi) energy increases.\cite{footnote1} The
successful growth and characterization of AlAs 2DESs
\cite{depoortereAPL02} with controllable valley occupation
\cite{shayeganPhysicaB06} opened up new opportunities to study the
effects of interaction in the presence of two discrete degrees of
freedom: spin and valley. The observation of a reduced $\chi_{s}$ in
a two-valley system compared to a single-valley case,
\cite{shkolnikovPRL04} was contrary to the then popular notion that
a two-valley system is effectively more dilute than its
single-valley counterpart due to its smaller Fermi energy. This
observation has been since explained theoretically to be the result
of a dominance of correlation effects.\cite{zhangPRB05} Recently,
valley susceptibility ($\chi_{v}$) measurements, where the response
of the system to an externally applied strain is studied, were
reported mainly for the case where both spins were
present.\cite{gunawanPRL06,footnote2} Here, we take this problem one
step further by measuring $\chi_{v}$ for the case when the system is
completely spin-polarized; we observe higher $\chi_{v}$ values
compared to the partially-polarized case. In other words, analogous
to the enhancement of $\chi_{s}$ upon complete valley-polarization,
we observe an enhancement of $\chi_{v}$ upon complete
spin-polarization. Our data suggest that the enhancement ensues
rather abruptly when the system moves from the partial to complete
spin-polarization regime.

We performed measurements on a 2DES confined to an 11 nm - thick
AlAs quantum well, grown using molecular beam epitaxy on a
semi-insulating GaAs (001) substrate. The AlAs well is flanked by
AlGaAs barriers and is modulation-doped with Si.
\cite{depoortereAPL02} We fabricated the sample using standard
photolithography techniques. Contacts were made by depositing GeAuNi
contacts and alloying in a reducing environment. Metallic gates were
deposited on front and back of the sample, allowing us to control
the 2D electron density (\textit{n}). We made measurements in a
$^{3}$He system with a base temperature of 0.3 K using standard
low-frequency lock-in techniques.

Bulk AlAs has three ellipsoidal conduction band minima (valleys) at
the X-points of the Brillouin zone. In wide AlAs quantum wells, the
biaxial compressive strain due to the slightly larger lattice
constant of AlAs compared to GaAs favors the occupation of the two
valleys with their major axes lying in the 2D plane.
\cite{shayeganPhysicaB06} We denote these valleys as \textit{X} and
\textit{Y}, according to the direction of their major axes [see
Fig.\ 1(a)]. They have anisotropic in-plane Fermi contours
characterized by transverse and longitudinal band effective masses,
\textit{m$_{t}$} = 0.205\textit{m$_{e}$} and \textit{m$_{l}$} =
1.05\textit{m$_{e}$}, where \textit{m$_{e}$} is the free electron
mass. This means that the relevant (density-of-states) band
effective mass in our 2DES is \textit{$m_{b}$} =
$\sqrt{\textit{m$_{t}$}\textit{m$_{l}$}}$ = 0.46\textit{m$_{e}$}.

\begin{figure*}[t]
\includegraphics[scale=1]{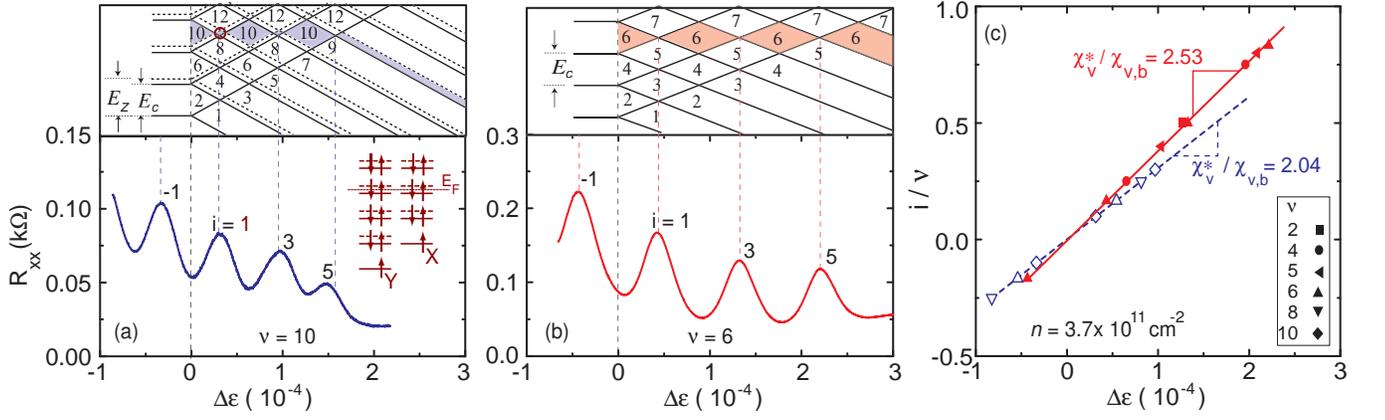}
\caption{ (color online) (a) Coincidence method of measuring
$\chi_{v}^{*}$ in the partially spin-polarized regime shown for $n$
= $3.7 \times 10^{11}$ cm$^{-2}$ at $\theta = 0^{\circ}$.
Corresponding energy diagram is shown in the top panel. The bottom
panel shows the piezoresistance trace for $\nu=10$ where the maxima
correspond to coincidences in the energy diagram. The inset shows
energy levels for the $i=1$ peak which is also marked by the open
brown circle in the top panel. (b) Similar measurements for the
fully spin-polarized case taken at a high angle of $\theta =
74^{\circ}$. (c) Index plot of data shown in the left and middle
panels, combined with similar measurements at other $\nu$'s. The
blue (open) and red (closed) symbols represent partial and complete
spin-polarizations. The slopes give $\chi_{v}^{*}$.}
\end{figure*}

Figure 1 shows how we achieve independent control over the valley
and spin degrees of freedom via the application of in-plane uniaxial
strain and external magnetic field respectively. The degeneracy
between the two in-plane valleys can be broken with controllable
strain, $\epsilon=\epsilon_{[100]}-\epsilon_{[010]}$ where
$\epsilon_{[100]}$ and $\epsilon_{[010]}$ denote strain along the
[100] and [010] crystal directions respectively.
\cite{shayeganPhysicaB06} To implement this, we glue the sample on a
piezo-electric actuator (piezo).
\cite{shayeganPhysicaB06} A voltage bias applied to
the piezo induces in-plane strain in the sample and causes a
transfer of electrons from one valley to the other as depicted in
Fig.\ 1(a). The induced valley splitting is given by $E_{v}=\epsilon
E_{2}$ where $E_{2}$ is the deformation potential which has a band
value of 5.8eV for AlAs. Analogous to its widely probed spin
counterpart, valley susceptibility is defined \cite{gunawanPRL06} as
$\chi_{v}=d\Delta n_{v}/d \epsilon$ where $\Delta n_{v}$ is the
difference in the electron population in $X$ and $Y$ valleys,
($n_{Y}-n_{X}$). In a non-interacting picture, we have,
$\chi_{v,b}\propto E_{2}m_{b}$. In a Fermi liquid picture, the
presence of interaction is accounted for by renormalized quantities,
denoted with asterisks throughout this paper. That is, in an
interacting system, we have $\chi_{v}^{*}\propto E_{2}^{*}m^{*}$.

In our study we probe the system under partial and complete
spin-polarizations. This is shown in Fig.\ 1(c) where the top
schematics show how a finite valley splitting is introduced in the
system when one (right) or two (left) spin species are occupied. In
Fig.\ 1(b) we show the experimental setup which is used to control
the level of spin-polarization. The sample is oriented at an angle
($\theta$) with respect to an external magnetic field so that it is
subjected to both perpendicular ($B_{\perp}$) and parallel
($B_{\parallel}$) components of the field. Magnetic field introduces
a Zeeman energy $E_{Z}=g\mu_{B}B_{tot}$ where $g$ is the Lande
$g-$factor and $\mu_{B}$ is the Bohr magneton. At high enough
$\theta$, $E_{Z}$ becomes greater than the Fermi energy ($E_{F}$)
and the system becomes completely spin-polarized.

Given that spin and valley are two discrete degrees of freedom, we
find it instructive to compare the measurements of $\chi_{v}^{*}$
and $\chi_{s}^{*}$ in the same system. The bottom schematics in
Fig.\ 1(c) show the spin splitting when the sample is subjected to a
magnetic field under single- (right) and two- (left) valley
occupation. We will return to our $\chi_{s}^{*}$ measurements later
in the paper.

In Fig.\ 2, we show the details of our valley susceptibility
measurements\cite{footnote3} for $n = 3.7 \times 10^{11}$cm$^{-2}$.
For the data of Fig.\ 2(a), the sample is held at a constant angle,
$\theta = 0^{\circ}$. The application of a magnetic field causes the
formation of Landau levels (LLs) which are split by cyclotron
energy, $E_{c}=\hbar eB_{\perp}/m^{*}$. The LLs of opposite spin are
further split by $E_{Z}$. The corresponding energy level diagram is
shown in the upper panel of Fig.\ 2(a). Note that, in our system,
for the density shown, $E_{Z}>E_{c}$ even for $\theta = 0^{\circ}$.
The LLs of one spin are shown as solid lines, while the dotted lines
denote LLs belonging to the opposite spin. Notice that when
$\Delta\epsilon = 0$ (see Ref.\ 17) each of these spin-split levels
is doubly valley-degenerate. We then apply an in-plane strain which
breaks this degeneracy and introduces a finite valley splitting. For
any given $\nu$, there are specific values of $\Delta\epsilon$ at
which the energy levels corresponding to the $X$ and $Y$ valleys
come into coincidence. \cite{gunawanPRL06} For example, as indicated
by the blue shaded region in the upper panel of Fig.\ 2(a), the
energy gap at LL filling factor $\nu$ = 10 oscillates as the applied
strain causes coincidences between the LLs, and finally saturates
after the system becomes completely valley-polarized. The lower
panel of Fig.\ 2(a) shows the corresponding piezoresistance trace.
Note that a large value of gap corresponds to a minimum in the trace
while a coincidence is marked by a peak. The small blue diamonds in
the top panel are unresolvable in our experiment. The oscillations
are periodic \cite{footnote4} and the positions of the peaks give a
measure of $\chi_{v}^{*}$. \cite{gunawanPRL06} For example, the
condition for coincidence at $\nu$ = 10 is $E_{v} = iE_{c}$ where
$i$ is an odd integer. This condition can also be written as
$E_{2}^{*}m^{*} = i\hbar eB_{\perp}/\Delta\epsilon$ or $i/\nu =
(E_{2}^{*}m^{*}) \times (2\pi/nh^{2})\times(\Delta\epsilon) $. In
Fig.\ 2(c) we plot $i/\nu$ vs. $\Delta\epsilon$ (denoted by blue
open points) for the data of Fig.\ 2(a). The periodicity of the
oscillations in Fig.\ 2(a) implies that the resulting $i/\nu$ vs.
$\Delta\epsilon$ plot is a straight line. Repeating the same
measurement for different $\nu$'s, we observe that all points fall
on the same blue (dashed) line, \cite{gunawanPRL06} the slope of
which gives a $\nu$-independent $\chi_{v}^{*}$.

Results of similar experiments done on a completely spin-polarized
system are shown in Fig.\ 2(b). The sample is tilted to a high tilt
angle ($\theta = 74^{\circ}$) so that $E_{Z}$ is larger than
$E_{F}$. The corresponding fan diagram is shown in the top panel.
Note how only one spin level is present. The variation of the energy
gap with strain is shown for $\nu=6$ as the red shaded region. The
oscillations observed in the piezoresistance shown in the bottom
panel are well-described by the simple fan diagram. The plot of
$i/\nu$ vs. $\Delta\epsilon$ is shown by closed red symbols in Fig.\
2(c) which also includes data from  similar measurements made at
$\nu$ = 4 and 5. The data points all fall on the same line and give
a value of $\chi_{v}^{*}$ which is higher than the partially
spin-polarized case.

\begin{figure}  \centering \includegraphics[scale=1] {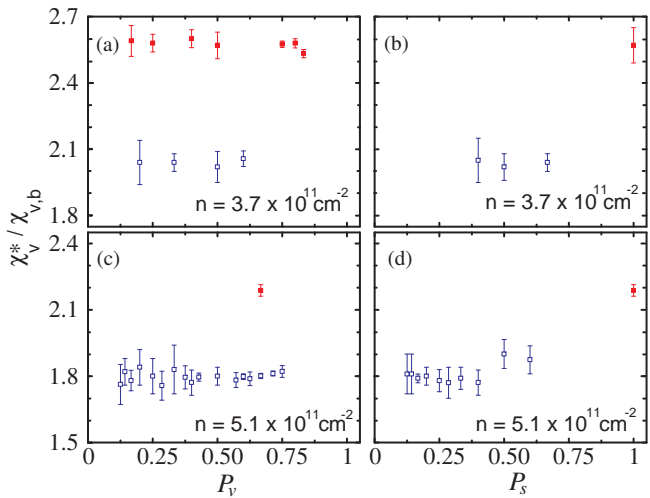}
\caption{ (color online) $\chi_{v}^{*}$, normalized to its band
value, as a function of $P_{v}$ and $P_{s}$ for two densities. The
blue (open) and red (closed) symbols correspond to partially and
completely spin-polarized regimes respectively. Panels (a) and (b)
are for $n = 3.7 \times 10^{11}$cm$^{-2}$ while (c) and (d)
correspond to $n = 5.1 \times 10^{11}$cm$^{-2}$. Each data point is
an average of values obtained for positive and negative values of
$\Delta\epsilon$, and for different tilt angles and fillings. }
\end{figure}

Figure 2(c) suggests that $\chi_{v}^{*}$ values divide themselves
into two groups corresponding to partial and complete
spin-polarizations. In each of the individual branches,
$\chi_{v}^{*}$ seems to be independent of the degree of valley and
spin polarizations. Valley-polarization of the system at any
particular $\nu$ is quantified as
$P_{v}=|(n_{Y}-n_{X})/(n_{Y}+n_{X})|$ where $n_{X}$ and $n_{Y}$
denote the occupation of the $X$ and $Y$ valleys. Similarly,
spin-polarization is defined as
$P_{s}=|(n_{\uparrow}-n_{\downarrow})/(n_{\uparrow}+n_{\downarrow})|$
where $n_{\uparrow}$ and $n_{\downarrow}$ denote the density of
electrons of up and down spins. Notice that, in general, each
coincidence in the bottom panels of Figs.\ 2(a) and (b) corresponds
to a different value of $P_{v}$ and $P_{s}$. As an example, in Fig.\
2(a) we show the energy levels for the $i=1$ coincidence of $\nu =
10$ as inset. In this case $P_{v}=0.2$ and $P_{s}=0.4$. In a similar
way, each of the maxima in the bottom panel of Fig.\ 2(b)
corresponds to a different value of $P_{v}$, e.g., the $i =$ 1, 3, 5
maxima correspond to $P_{v} =$ 16$\%$, 50$\%$ and 83$\% $,
respectively.

To bring out the dependencies of valley susceptibility, it is
instructive to plot it as an explicit function of $P_{v}$ and
$P_{s}$. Note that each data point in Fig. 2(c), in general,
corresponds to a different value of $P_{v}$ and $P_{s}$. From each
of these points we draw a line to the origin and use the slope to
determine the corresponding $\chi_{v}^{*}$. In Fig.\ 3(a), these
$\chi_{v}^{*}$ values, averaged over positive and negative
$\Delta\epsilon$, are shown as a function of $P_{v}$. The blue
(open) and red (closed) symbols in Fig. 3(a) represent the partially
and completely spin-polarized regimes, respectively.\cite{footnote5}
The division of points into two branches corresponding to the two
spin-polarization regimes is clear. Within each of these branches,
we do not observe dependence on $P_{v}$. The effect of complete
spin-polarization becomes more apparent in Fig.\ 3(b) where we plot
the same points as a function of $P_{s}$. $\chi_{v}^{*}$ is
independent of $P_{s}$ when $0.40 < P_{s} < 0.67$ but increases when
the system becomes completely spin-polarized.

We repeated these measurements for various densities where larger
ranges of $P_{v}$ and $P_{s}$ were accessible. Data for $n = 5.1
\times 10^{11}$cm$^{-2}$  are shown in Figs.\ 3(c) and (d). In the
partially spin-polarized regime, for $0.12 < P_{v} < 0.75 $ and
$0.12 < P_{s} < 0.6$, we observe an almost constant $\chi_{v}^{*}$
with a very weak increase at the higher end of $P_{s}$. In the
completely spin-polarized regime, we measure a much higher value of
$\chi_{v}^{*}$.

Summarizing all data from different densities, we conclude that
$\chi_{v}^{*}$ increases rather abruptly as the system makes a
transition from partial to complete spin polarization.
$\chi_{v}^{*}$ is otherwise largely independent of $P_{v}$ or
$P_{s}$ in either of these regimes.

Enhancement of spin-susceptibility when valley degree of freedom is
frozen out has been observed before. \cite{shkolnikovPRL04} Spin
susceptibility is defined as $\chi_{s}=d\Delta n_{s}/dB$ where
$\Delta n_{s}$ represents the imbalance in spin population,
($n_{\uparrow}-n_{\downarrow}$). We measured $\chi_{s}^{*}$ in our
sample using the widely used coincidence technique as the tilt angle
$\theta$ is varied. \cite{shkolnikovPRL04} The results of these
measurements are shown side by side with $\chi_{v}^{*}$, along with
the relevant polarization ranges, in Fig.\ 4. The left panel shows
our $\chi_{v}^{*}$ measurements: the blue (open) and red (closed)
symbols are for the partially and completely spin-polarized systems,
respectively. Special care has been taken to make sure that, for a
given $n$, $P_{v}$ is held constant while $P_{s}$ is changing from
partial to complete, so that the transition from the bottom to the
top branch is brought about by complete spin-polarization. The right
panel of Fig.\ 4 shows $\chi_{s}^{*}$ measurements for single-
(closed red symbols) and two- (open blue symbols) valley cases.
Again, for a given $n$, $P_{s}$ is held at similar values for the
top and bottom branches. We can see from this figure that
$\chi_{s}^{*}$ is higher for the single-valley case compared to the
two-valley case, consistent with earlier studies.
\cite{shkolnikovPRL04}

\begin{figure} \includegraphics[scale=1]{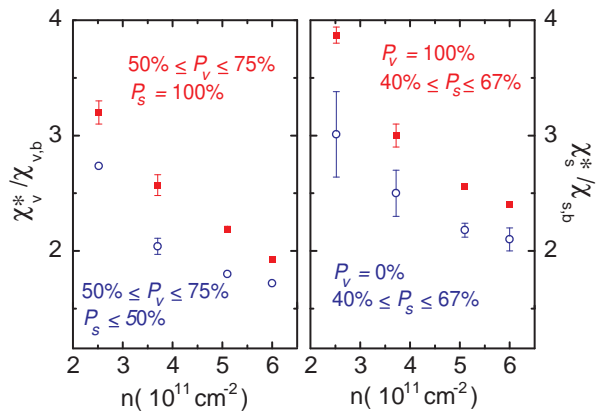}
\caption{(color online). Summary of susceptibility data. The left
panel shows $\chi_{v}^{*}$ for partial (blue circles) and complete
(red squares) spin-polarization. $P_{v}$ is held constant within a
narrow range to facilitate comparison. The right panel shows
$\chi_{s}^{*}$ measurements for zero (blue circles) and complete
(red squares) valley-polarizations. Again, $P_{s}$ is held constant
within a narrow range to minimize its influence on $\chi_{s}^{*}$.
Both $\chi_{v}^{*}$ and $\chi_{s}^{*}$ are normalized to their
respective band values. }
\end{figure}

The most notable feature in Fig.\ 4 is that all susceptibilities are
increasingly enhanced over their respective band values as $n$ is
decreased, as expected in an interacting electron picture. Another
remarkable feature is the similarity of the numerical values of
$\chi_{v}^{*}$ and $\chi_{s}^{*}$ in spite of the fact that they
represent the system's response to very different external stimuli.
This observation strongly suggests a parallel between spin and
valley as two discreet and independent degrees of freedom. It is
noteworthy though, that the values of $\chi_{v}^{*}$ and
$\chi_{s}^{*}$ are not exactly the same, possibly pointing towards subtle differences.

In an earlier study, \cite{gokmenPRB07} in a wider 2DES, a weak
dependence of $\chi_{s}^{*}$ on $P_{s}$ was reported in a
single-valley system.
 For $n = 5.5 \times 10^{11}$cm$^{-2}$ and for $0.29 \leq P_{s} \leq 0.50$, a $7\%$ variation was reported. As argued in Ref.\ 21, such
 dependence is reasonable and consistent with the large width (15nm) of the AlAs quantum well used. Since the well width of our sample is
 less than that used in Ref.\ 21, we expect a less prominent effect. Consistent with this expectation, for
 $n = 6.0 \times 10^{11}$cm$^{-2}$ and $0.15 \leq P_{s} \leq 0.40$, we observe a $2.5\%$ variation of $\chi_{s}^{*}$ in a single-valley system.
 However, it is noteworthy that in our studies of $\chi_{v}^{*}$ in a single-spin system, we do not find evidence for dependence on $P_{v}$.
 This is perhaps another indication of the inequality of spin and valley degrees of freedom. The fact that the orbital wavefunction for the
 two spins are the same, but that this is not the case for the two valleys, could conceivably alter the effect of interaction.

In summary, we measured valley susceptibility in an AlAs 2DES as a
function of both spin and valley polarizations. We observe that the
value of $\chi_{v}^{*}$ undergoes a rather sudden increase as the
system moves from partial to complete spin-polarization. Apart from
this, $\chi_{v}^{*}$ is mostly independent of $P_{s}$ and $P_{v}$.
We also measured the spin susceptibility for valley-polarized and
unpolarized systems. $\chi_{s}^{*}$ is observed to be higher in a
single-valley case compared to its two-valley counterpart. All
susceptibilities increase as $n$ is decreased, consistent with
increasing interaction.

We thank the NSF for support.


\begin{references}
\bibitem{footnote2} Measurements of Ref. 15 focused mainly on partially
spin-polarized 2DESs. Also reported was a single data point for the
fully spin-polarized case, which fell slightly above the
partially-polarized data points. Not enough data points were
measured, however, to reach a general conclusion.
\bibitem{footnote4} The last oscillation in this trace does not have the same period as others. However, this is not observed in other samples and hence we ignore the last oscillation in our susceptibility determinations.
\bibitem{footnote5} In Fig. 3(a), note that the average of the values indicated
by the blue (open) symbols is very close to the value of
$\chi_{v}^{*}$ derived from the slope indicated in Fig. 2(c). This
is not surprising, given the linearity of the points in Fig. 2(c).
\end{references}
\end{document}